\documentclass[conference]{IEEEtran}

\usepackage{amssymb,stmaryrd,amsmath,amsfonts,rotating,mathrsfs}%amsfonts,amsthm
\usepackage[noadjust]{cite}
\usepackage{color}%temporary packages
\usepackage[vflt]{floatflt}
\usepackage{epic}
\definecolor{TODO}{rgb}{0.6,0.6,0.6} % TO DO!!!!

\definecolor{TOCHECK}{rgb}{0.8,0.8,0.8} % TO CKECK!!!!

\newtheorem{theorem}{Theorem}

\newcommand{\btheo}{\begin{theorem}}
\newcommand{\etheo}{\end{theorem}}
\newcommand{\bproof}{\begin{proof}}
\newcommand{\eproof}{\end{proof}}
\newtheorem{definition}[theorem]{Definition}
\newcommand{\bdefi}{\begin{definition}}
\newcommand{\edefi}{\end{definition}}
\newtheorem{fact}[theorem]{Fact}
\newcommand{\bprop}{\begin{fact}}
\newcommand{\eprop}{\end{fact}}
\newtheorem{corollary}[theorem]{Corollary}
\newcommand{\bcor}{\begin{corollary}}
\newcommand{\ecor}{\end{corollary}}
\newtheorem{example}[theorem]{Example}
\newcommand{\bex}{\begin{example}}
\newcommand{\eex}{\end{example}}
\newtheorem{lemma}[theorem]{Lemma}
\newcommand{\blemma}{\begin{lemma}}
\newcommand{\elemma}{\end{lemma}}
\newtheorem{remark}[theorem]{Remark}
\newcommand{\bremark}{\begin{remark}}
\newcommand{\eremark}{\end{remark}}
\newtheorem{conj}[theorem]{Conjecture}
\newcommand{\bconj}{\begin{conj}}
\newcommand{\econj}{\end{conj}}

\newcommand{\naturals}{\ensuremath{\mathbb{N}}}

\def\0{{\tt 0}} % Ex.: BPSK modulation => 0 is encoded into +1
\def\1{{\tt 1}} % Ex.: BPSK modulation => 1 is encoded into -1
\def\?{{\tt *}} % erasure symbol
\renewcommand{\mid}{\,|\,}

\newcommand{\EEx}{\hfill $\Diamond$}

\RequirePackage{bbm}
\allowdisplaybreaks

\ifCLASSINFOpdf

\else

\fi

\begin{document}
\title{Polar Codes: Robustness of the Successive Cancellation Decoder with Respect to Quantization}

\author{\IEEEauthorblockN{S. Hamed Hassani and R\"udiger Urbanke}
\IEEEauthorblockA{School of Computer and Communication Sciences, 
EPFL\\
Email: \{seyedhamed.hassani, rudiger.urbanke\}@epfl.ch}
% \and
% \IEEEauthorblockN{Homer Simpson}
% \IEEEauthorblockA{Twentieth Century Fox\\
% Email: homer@thesimpsons.com}
% \and
% \IEEEauthorblockN{James Kirk\\ and Montgomery Scott}
% \IEEEauthorblockA{Starfleet Academy\\
% San Francisco, California 96678-2391\\
% Telephone: (800) 555--1212\\
% Fax: (888) 555--1212}
}


%Telephone: (800) 555--1212, Fax: (888) 555--1212}


\maketitle

\begin{abstract}
%\boldmath
Polar codes provably achieve the capacity of a wide array of channels
under successive decoding. This assumes infinite precision arithmetic.
Given the successive nature of the decoding algorithm, one might
worry about the sensitivity of the performance to the
precision of the computation.

We show that even very coarsely quantized decoding algorithms 
lead to excellent performance. More concretely, we show that under
successive decoding with an alphabet of cardinality only three, the
decoder still has a threshold and this threshold is a sizable
fraction of capacity. More generally, we show that if we are willing
to transmit at a rate $\delta$ below capacity, then we
need only $c \log(1/\delta)$ bits of precision, where $c$ is a
universal constant.
\end{abstract}


% \fi
\IEEEpeerreviewmaketitle

\section{Introduction}
Since the invention of polar codes by Arikan, \cite{Ari09}, a large body of
work has been done to investigate the pros and cons of  polar
codes in different practical scenarios (for a partial list see \cite{MT09}-\cite{HKU09}).

We address one further aspect of polar codes using successive
decoding. We ask whether such a coding scheme is {\em robust}. More
precisely, the standard analysis of polar codes under successive
decoding assumes infinite precision arithmetic. Given the successive
nature of the decoder, one might worry how well such a scheme performs
under a finite precision decoder. A priori it is not clear whether
such a coding scheme still shows any threshold behavior and, even
if it does, how the threshold scales in the number of bits of the
decoder.

\begin{figure}[h]
\begin{centering}
\input{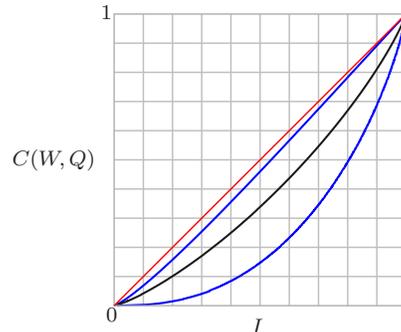}
\caption{\small 
The maximum achievable rate, call it $C(W,Q)$, of a simple three message decoder, called the decoder with erasures, 
as a function of the capacity of the channel for different channel families.
From top to bottom: the first curve corresponds to the family of binary erasure channels (BEC) where the decoder with erasures is equivalent to
the original SC decoder and, hence, the maximum achievable rate is  the capacity itself. The second curve corresponds to the 
family of binary symmetric channels (BSC). The third curve corresponds to the family of binary additive white Gaussian channels 
(BAWGN). The curve at the bottom corresponds to a universal lower bound on the achievable rate by the decoder with erasures.   \label{fig:dwecap}
}
\end{centering}
\end{figure}

We show that in fact polar coding is extremely robust with respect
to the quantization of the decoder. In Figure~\ref{fig:dwecap},
we show the achievable rate using a simple successive decoder
with only three messages, called the decoder with erasures, when transmission takes place over several important channel 
families. As one can see from this figure, in particular for channels with high capacity,
the fraction of the capacity that is achieved by this simple decoder is close to $1$, i.e., even this extremely simple
decoder almost achieves capacity. We further show that, more
generally, if we want to achieve a rate which is $\delta$ below capacity by $\delta>0$,
then we need at most  $c\log(1/\delta)$ bits of precision (all the logarithms in this paper are in base 2).

The significance of our observations goes beyond the pure computational
complexity which is required. Typically, the main bottleneck in the implementation
of large high speed coding systems is memory. Therefore,
if one can find decoders which work with only a few bits per message
then this can make the difference whether a coding scheme
is implementable or not.

\subsection{Basic setting and definitions}
Let $W: \mathcal{X} \to \mathcal{Y}$ be a binary memoryless symmetric (BMS) channel, 
 with input alphabet $\mathcal{X}
=\{0,1\}$,  output alphabet  $\mathcal{Y}$, and the 
transition probabilities $\{W(y \mid x): x \in \mathcal{X}, y \in \mathcal{Y}\}$. 
Also, let $I(W)$ denote the capacity of $W$.

Let $G_2=\left[\begin{smallmatrix}1&0\\1&1\end{smallmatrix}\right]$.
The generator matrix of polar codes is defined through the Kronecker
powers of $G_2$, denoted by $G_N=G_2^{\otimes n}$.  
Throughout the paper, the variables $N$ and $n$
are related as $N=2^n$.
Let us 
review very briefly how the generator matrix of polar codes is constructed. 
Consider the $N \times N$ matrix $G_N$ and let us label the rows
of the matrix $G_N$ from top to bottom by $0, 1, \cdots, N-1$. Now
assume that we desire to transmit binary data over the channel $W$
at rate $R < I(W)$ with block-length $N$.  One way to accomplish this
 is to choose a subset $\mathcal{I} \subseteq \{0, \cdots, N-1\}$ of
size $NR$ and to construct a vector $U_0^{N-1}=(U_0, \cdots, U_{N-1})$
in a way that it contains our $NR$ bits of data at positions in $\mathcal{I}$
and contains, at positions not in $\mathcal{I}$,  some fixed value (for example
$0$) which is known to both the encoder and decoder. We then send
the codeword $X_0^{N-1}= U_0^{N-1}G_N$ through the channel $W$.  We
refer to the set $\mathcal{I}$ as the set of {\em chosen indices} or  {\em
information indices} and the  set $\mathcal{I}^c$ is called the set of
{{\em frozen indices}}. We explain in Section~\ref{sec:tree} how 
the good indices are chosen. 
 At the decoder, the bits $u_0, \cdots, u_{N-1}$ are decoded one by one. That is, the bit $u_i$ is decoded after $u_0, \cdots u_{i-1}$. 
If $i$ is a frozen index, its value is known to the decoder. If not,  the decoder estimates the value of $u_i$
 by using the output $y_0^{N-1}$ and the estimates of $u_0, \cdots, u_{i-1}$.

\subsection{Quantized SC decoder}\label{sec:equivalentmodel}
Let $\mathbb{R}^* = \mathbb{R} \cup \{ \pm \infty \}$ and consider
a function $Q(x): \mathbb{R}^* \to \mathbb{R}^*$ that is {\em
anti-symmetric } (i.e., $Q(x)= -Q(-x)$).  We define the $Q$-quantized SC decoder as a version of
the SC decoder
in which the function $Q$ is applied to the output of any computation
that the SC decoder does.  We denote such a decoder by $\text{SCD}_{Q}$.  

Typically, the purpose of the function $Q$ is to model the 
case where we only have finite precision in our computations perhaps due to 
 limited available memory or due to other hardware 
limitations. Hence, the computations are correct within a 
certain level of accuracy which the function $Q$  models.  Thus,  let us assume that the 
range of $Q$ is a finite set $\mathcal{Q}$ with cardinality $\mid \mathcal{Q} \mid$.  As a result, all the 
messages passed through the decoder $\text{SCD}_Q$ belong to the set $\mathcal{Q}$.

In this paper we consider a simple choice of the function $Q$ that is
 specified by two parameters: The distance between levels $\Delta$, 
and truncation threshold $M$. 
Given a specific choice of $M$ and $\Delta$, we define $Q$ as follows: 
\begin{equation} \label{Q}
Q (x)=  \left\{
\begin{array}{lr}
\big \lfloor \frac{x}{\Delta} + \frac 12 \rfloor \Delta, &   \quad x \in [-M,M],\\ \\
\text{sign}(x)M,  & \quad \text{otherwise}.
\end{array} \right.
\end{equation}
Note here that $\mid \mathcal{Q} \mid= 1+\frac{2M}{\Delta}$.

\subsection{Summary of results} \label{sec:mainresult}
\begin{theorem}[Main Statement] \label{main}
Consider transmission over a BMS channel $W$ of capacity $I(W)$ using polar codes and a $\text{SCD}_{Q}$
with message alphabet $Q$. Let $C(W, Q)$ denote the maximum rate at which
reliable transmission is possible for this setup.
\begin{itemize}
\item[(i)] Let $|{\mathcal Q}|=3$. Then there exists a computable decreasing sequence $\{U_n\}_{n \in \naturals}$ (see  \eqref{U_n}) and
a computable increasing sequence $\{L_n\}_{n \in \naturals}$ (see \eqref{L_n}), 
so that $L_n \leq C(W, Q) \leq U_n$ and
\begin{align*}
\lim_{n \rightarrow \infty} L_n =
\lim_{n \rightarrow \infty} U_n.
\end{align*}
In other words, $U_n$ is an upper bound and $L_n$ is a lower bound
on the maximum achievable rate $C(W, Q)$ and for increasing $n$ these two bounds converge to $C(W,Q)$.  

\item[(ii)] To achieve an additive gap $\delta >0$ to capacity $I(W)$, it suffices to choose 
$\log |{\mathcal Q}| = c \log(1/\delta)$. \QED
\end{itemize}
\end{theorem}
\emph{Discussion:} In Figure~\ref{fig:dwecap} the value of $C(W,Q)$, $|\mathcal{Q}|=3$, 
is plotted  as a function of $I(W)$ for different channel families (for more details see Section~\ref{compute}). 
A universal lower bound for 
the maximum achievable rate is also given in Figure~\ref{fig:dwecap}. 
This suggests that even for small values of $|\mathcal{Q}|$ polar codes are very robust to quantization. In particular for channels with 
capacity close to $1$, very little is lost by quantizing.
The methods used here are extendable to other quantized decoders.

The rest of the paper is  devoted to proving the first part of Theorem~\ref{main}. 
Due to space limitation, we have omitted the proof of the second part of theorem~\ref{main} as well as the proofs 
of the lemmas stated in the sequel 
and we refer the reader to \cite{long} for more details.

\section{General Framework for the Analysis}\label{sec:analysis}

\subsection{Equivalent tree channel model and analysis of the probability 
of error for the original SC decoder} \label{sec:tree}
 Since we are dealing with a linear code, 
a symmetric channel
 and symmetric 
decoders throughout this paper, without loss of 
generality we confine ourselves to the \emph{all-zero codeword} 
(i.e., we assume that all the $u_i$'s are equal to $0$).
% \footnote{In 
% terms of the analysis of the probability of error, it 
% must be noted that since the we are dealing with a 
% symmetric channel and a symmetric decoder, for any
%  codeword the average error probability is the 
% same as the average error probability for the  
% all-zero error codeword  (\cite[Chapter 4]{RiU08})}. 
In order to better visualize the decoding process, the following definition is handy.    

\begin{definition}[Tree Channels of Height $n$]
For each $i \in \{0,1, \cdots, N-1 \}$, we introduce the notion of the $i$-th tree channel of height $n$ which is denoted by $T(i)$. Let
$b_{1}\dots b_n$ be the $n$-bit binary expansion of
$i$. E.g., we have for $n=3$, $0=000$, $1=001$, \dots, $7=111$. With a slight abuse of notation we use 
$i$ and $b_1 \cdots b_n$ interchangeably. Note
that for our purpose it is slightly more convenient to denote the least
(most) significant bit as $b_n$ ($b_1$).  Each tree
channel consists of $n+1$ levels, namely $0,\dots,n$. It is a complete
binary tree. The root is at level $n$. At level $j$ we have $2^{n-j}$
nodes. For $1 \leq j \leq n$, if $b_{j} = 0$ then all nodes on level
$j$ are check nodes; if $b_{j} = 1$ then all nodes on level $j$
are variable nodes. Finally, we give a label for each node in the tree $T(i)$: For each level $j$, we label the $2^{n-j}$ nodes at this level respectively from left to right by
$(j,0), (j,1), \cdots, (j,2^{n-j}-1)$.

  All nodes at level $0$ correspond to independent
observations of the output of the channel $W$, assuming that the input
is $0$.

An example for $T(3)$ (that is $n=3$, $b=011$ and $i=3$) is shown in
Fig.~\ref{fig:tree}.
\begin{figure}[!h] \begin{center} \input{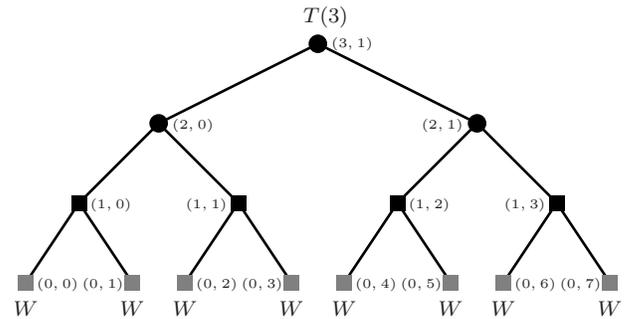} \end{center}
\caption{ Tree representation of the tree-channel $T(3)$. The $3$-bit binary
expansion of $3$ is $b_1b_2 b_3 = 011$ (note that $b_1$ is the most significant bit). The pair beside each node is the label 
assigned to it.\label{fig:tree}}
\end{figure}

\end{definition}

Given the channel output vector $y_0^{N-1}$ and assuming that the values of the bits 
prior to $u_i$ are given,  i.e., 
$u_0=0, \cdots, u_{i-1}=0$, we now  compute the probabilities 
$p(y_0^{N-1}, u_0^{i-1} \mid u_i=0)$ and
$p(y_0^{N-1}, u_0^{i-1} \mid u_i=1)$ via a simple message passing procedure
on the equivalent  tree channel $T(i)$. We attach to each node 
in $T(i)$ with label $(j,k)$ a 
message\footnote{To simplify notation, we drop the dependency 
of the messages $m_{j,k}$ to the position $i$ whenever it is clear from the context.}
 $m_{j,k}$ and we update the messages as we go up towards the root node. We start with initializing the messages at the leaf nodes of $T(i)$.
For this purpose, it is convenient to represent the channel
in the log-likelihood domain; i.e., for the node with label $(0,k)$ at the bottom of the tree which corresponds to an
independent realization of $W$, we plug in the log-likelihood ratio (llr) $\log(\frac{W(y_k\mid 0)}{W(y_k\mid 1)})$ 
as the initial message $m_{0,k}$. That is,
\begin{equation}
m_{0,k}= \log(\frac{W(y_k\mid 0)}{W(y_k\mid 1)}).
\end{equation}

Next, the SC decoder recursively computes 
the messages (llr's) at each level via the following operations:  If the nodes at level $j$ are variable nodes (i.e., $b_j=1$), we have

 \begin{equation} \label{var_u}
m_{j,k}=m_{j-1,2k}+ m_{j-1,2k+1},
\end{equation}
and if the nodes at level $j$ are check nodes (i.e., $b_j=0$), the message that is passed up is
\begin{equation} \label{check_u}
 m_{j,k}=2 \tanh^{-1}(\tanh(\frac{m_{j-1,2k}}{2}) \tanh(\frac{m_{j-1,2k+1}}{2})).
\end{equation}
%For future reference, let us define the variable update function $v(x,y)$ and the check update function $c(x,y)$ as
%\begin{align}
%& v(x,y)=x+y, \\
%& c(x,y)=2 \tanh^{-1}(\tanh{(\frac{x}{2}}) \tanh({\frac{y}{2}})).
%\end{align} 
%Hence, the message mounting from a variable (resp. check) node will be the result of applying the function $v(x,y)$ (resp. $c(x,y)$) on its two incoming messages.
 In this way, it can be shown that (\cite{Ari09}) the message that we obtain at the root node is precisely the value
 \begin{equation}
m_{n,0}=\log(\frac{p(y_0^{N-1}, u_0^{i-1} \mid u_i=0)}{p(y_0^{N-1}, u_0^{i-1} \mid u_i=1)}).
 \end{equation}
Now, given $(y_0^{N-1},u_0^{i-1}) $, the value of $u_i$ is estimated as follows. If
$m_{n,0}>0$ we let $u_i=0$. If $m_{n,0}<0$ we let $u_i=1$. Finally, if $m_{n,0}=0$ we choose the value of
$u_i$ to be either $0$ or $1$ with probability $\frac 12$. 
Thus, denoting $E_i$ as the event that we make
 an error on the $i$-th bit within the above setting, we obtain
\begin{equation}
\text{Pr}(E_i) = \text{Pr}(m_{n,0} <0)+ \frac 12 \text{Pr} (m_{n,0}=0).
\end{equation}

Given the description of $m_{n,0}$ in terms of a tree channel,
it is now clear that we can use density evolution \cite{MT09} to compute 
the probability density function of $m_{n,0}$. In this regard, at each level $j$, the
 random variables $m_{j,k}$ are i.i.d. for $k \in \{0, 1, \cdots, 2^{n-j}-1\}$. The
 distribution of the 
leaf messages $m_{0,k}$  is the distribution of the variable
$\log(\frac{W(Y\mid 0)}{W(Y\mid 1)})$, where $Y\sim W(y\mid 0)$. One can recursively compute the distribution of 
$m_{j,k}$ in terms of the distribution of $m_{j-1,2k}, m_{j-1,2k+1}$ and the type of 
the nodes at level $j$ (variable or check) by using the relations  \eqref{var_u}, \eqref{check_u} with 
the fact that the random variables $m_{j-1,2k}$ and $m_{j-1,2k+1}$ are i.i.d. 

\subsection{Quantized density evolution}
Let us now analyze the density evolution procedure for the quantized decoder. For each label $(j,k)$ in $T(i)$, let  $\hat{m}_{j,k}$ 
represent the messages at this label. The messages $\hat{m}_{j,k}$ take their values 
in the discrete set $\mathcal{Q}$ (range of the function $Q$). It is now easy to see that 
for the decoder $\text{SCD}_Q$ the messages evolve via the following relations.
At 
the leaf nodes of the tree we plug in the message $\hat{m}_{0,k}= Q(\log(\frac{W(y_k\mid 0)}{W(y_k\mid 1)}))$,
and the update equation for $\hat{m}_{(j,k)}$ is 
 \begin{equation} \label{Qvar_u}
\hat{m}_{j,k}=Q(\hat{m}_{j-1,2k}+ \hat{m}_{j-1,2k+1}),
\end{equation}
if the node $(j,k)$ is a variable node and
\begin{equation} \label{Qcheck_u}
 \hat{m}_{j,k}=Q(2 \tanh^{-1}(\tanh(\frac{\hat{m}_{j-1,2k}}{2}) \tanh(\frac{\hat{m}_{j-1,2k+1}}{2}))),
\end{equation}
if the node $(j,k)$ is a check node. One can use the density evolution procedure 
to recursively obtain the densities of the messages $\hat{m}_{j,k}$. 

Finally, let $\hat{E}_i$ denote the event that we make an error in decoding the $i$-th bit, with a further assumption
that  we have correctly decoded the previous bits $u_0, \cdots, u_{i-1}$. In a similar way
 as in the analysis of the original SC decoder, we get
\begin{equation}
\text{Pr}(\hat{E}_i) = \text{Pr}(\hat{m}_{n,0} <0)+ \frac 12 \text{Pr} (\hat{m}_{n,0}=0).
\end{equation}
Hence, one way to choose the information bits for the algorithm $\text{SCD}_Q$ is
 to choose the bits $u_i$ according to the least values of 
$\text{Pr}(\hat{E}_i)$. 

An important point to note here is that with the decoder $\text{SCD}_{Q}$,
the distribution of the messages in the trees $T(i)$ is
 different than the corresponding ones that result from the
original SC decoder. Hence, the choice of the information indices
is also specified by the choice of the function $Q$ as well as the
channel $W$. 

Note here that, since all of the densities takes their value in the
 finite alphabet $\mathcal{Q}$, the construction of such polar 
codes can be efficiently done in time $O(\mid \mathcal{Q} \mid ^2 N \log N)$. We refer the reader to \cite{Ari09} for more details.

\subsection{Gallager Algorithm}
Since our aim is to show that polar codes under successive decoding
are robust against  quantization, let us investigate an extreme
case. The perhaps simplest message-passing type decoder one can
envision is the Gallager algorithm. It works with single-bit messages.
Does this simple decoder have a non-zero threshold? Unfortunately
it does not, and this is easy to see.
We start with the equivalent tree-channel model. Consider an arbitrary tree-channel $T(i)$. 
Since messages are only a single bit, the ``state'' of the decoder
at level $j$ of $T(i)$ can be described by a  single non-negative number,
namely the probability that the message at level $j$ is incorrect.
It is an easy exercise to show that at a level with check nodes the state becomes worse 
and at a level with variable nodes the state stays
unchanged and hence no progress in the decoding is achieved, irrespective
of the given tree.  In other words, this decoder has a threshold
of zero.  The problem is the processing at the
variable nodes since no progress is achieved there. But since we
only have two possible incoming messages there is not much degree of freedom
in the processing rules. 
%It is doubtful if any message-passing
%decoder with only a single-bit message can do better. 

\subsection{1-Bit Decoder with Erasures}
Motivated by the previous example, let us now add one message to the alphabet of the Gallager 
decoder, i.e.,  we also add the possibility of having erasures. In this case  $Q(x)$ becomes the
sign function\footnote{Note here that we have further assumed that $M=\Delta$ and $\Delta\to 0$.}, i.e.,
\begin{equation} \label{Qgal}
Q(x)=  \left\{
 \begin{array}{lr}
\infty, &    x>0,\\ 0, &  x=0, \\  \!\!\!\!   -\infty,
&    x<0. \\
 \end{array} \right.  \end{equation}

As a result, all messages passed by the algorithm $\text{SCD}_Q$
take on only three possible values: $\{ -\infty, 0, \infty \}$. In
this regard, the decoding procedure takes a very simple form. The
algorithm starts by quantizing the channel output to one of the
three values in the set   $\mathcal{Q}=\{ -\infty, 0, \infty \}$.  At a check node we
take the product of the signs of the incoming messages and at a
variable node we have the natural addition rule ($0  \leftarrow
\infty + -\infty$, $0 \leftarrow 0+0 $ and $\infty \leftarrow \infty+
\infty, \infty \leftarrow \infty+ 0 $ and $-\infty \leftarrow
-\infty+ -\infty, -\infty \leftarrow -\infty+ 0$ ). Note that on
the binary erasure channel, this algorithm is equivalent to the
original SC decoder.

Our objective is now to compute the maximum reliable rate that the decoder 
$\text{SCD}_Q$ can achieve for a BMS channel $W$.  We denote this quantity by $C(W,Q)$.  The analysis is done in three steps:
\subsubsection{The density evolution procedure} \label{1e}
To analyze the performance of this algorithm, first note that since all our messages take their values in the set $\mathcal{Q}$, then
all the random variables that we consider  have the following form
\begin{equation} \label{gal_den}
D=  \left\{
 \begin{array}{lr}
\infty, &   \text{w.p. $ p$},\\ 0, &   \text{w.p. $e$}, \\   \!\!\!\!  -\infty,
&   \; \;\;\;\;  \;\;\; \;\;\;\;  \;\;\;\;    \text{w.p. $m$}. \\
 \end{array} \right.  \end{equation}
Here, the numbers $p,e,m$ are probability values and
$p+e+m=1$. 
Let us now see how the density evolves through the tree-channels. 
For this purpose, one should trace the output distribution of \eqref{Qvar_u} and \eqref{Qcheck_u} when the input messages are two i.i.d. 
copies of a r.v. $D$ with pdf as in \eqref{gal_den}.  
\begin{lemma}
Given two i.i.d. versions of a r.v. $D$ with distribution as in \eqref{gal_den}, the output
of a variable node operation \eqref{Qvar_u}, denoted by $D^+$, has the
following form
% \begin{equation} \label{gal_plus}
% D^+= (m^2+ 2em)\Delta_{-\infty} + (e^2+ 2pm) \Delta_0 + (p^2+ 2pe)
% \Delta_{\infty}.
%\end{equation}   
\begin{equation}  
D^+=  \left\{   
\begin{array}{lr}
\label{gal_plus}  
\infty, &   \text{w.p. $ p^2+2pe$},\\  0, &  
\text{w.p. $e^2+2pm$}, \\  -\infty, &   \text{w.p. $m^2+2em$}. \\
\end{array} \right.    
\end{equation}
Also,  the check node operation \eqref{Qcheck_u},  yields $D^-$ as
%\begin{equation} \label{gal_min}
% D^-=  (m^2+ 2pm)\Delta_{-\infty} + (1-(1-e)^2) \Delta_0 + (p^2+
% m^2) \Delta_{\infty}.
%\end{equation}   
\begin{equation}  
D^-=  \left\{   
\begin{array}{lr}
\label{gal_min}  
\infty, &   \text{w.p. $ p^2+m^2$},\\  0, &  
\text{w.p. $1-(1-e)^2$}, \\  
\!\!\!\!-\infty, &   \text{w.p. $2pm$}. \\ 
\end{array} \right.    
\end{equation} 
\end{lemma}

In order to compute the distribution of the messages $\hat{m}_{n,0}$ at a given level 
$n$, we use the method of \cite{Ari09} and define the polarization process $D_n$
as follows. Consider the random variable  
$L(Y)=\log(\frac{W(Y\mid 0)}{W(Y\mid 1)})$, where $Y\sim W(y\mid 0)$. 
The stochastic
process $D_n$  starts from the r.v. $D_0=Q(L(Y))$ defined as
\begin{equation}   \label{D_0}
D_0=  \left\{   
\begin{array}{lr}
\infty, &   \text{w.p. $ p=\text{Pr}(L(Y) > 0)$},\\  0, &  
\text{w.p. $e=\text{Pr}(L(Y) = 0)$}, \\  
-\infty, &   \text{w.p.
$m=\text{Pr}(L(Y) < 0)$}, \\   
\end{array} \right.    
\end{equation}
and for $n \geq 0$ 
\begin{equation} D_{n+1}=  \left\{ \begin{array}{lr}
D_n ^{+}, &  \text{w.p. $\frac 12$},\\D_n ^{-}, &   \text{w.p.
$\frac 12$}, \end{array} \right.  \end{equation} where the plus and
minus operations are given in \eqref{gal_plus}, \eqref{gal_min}.
\subsubsection{Analysis of the process $D_n$} \label{compute} 
\begin{figure}[h]
\begin{centering}
\input{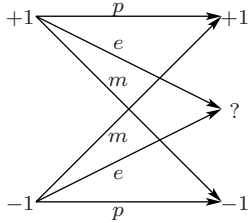}
\caption{\small 
The equivalent channel for the density $D$ given in \eqref{gal_den}. \label{fig:channel}
}
\end{centering}
\end{figure}

Note that the output of process $D_n$ is  itself a random variable
of the form given in \eqref{gal_den}. Hence, we can equivalently
represent the process $D_n$ with a triple $(m_n, e_n, p_n)$, where
the coupled processes $m_n, e_n$ and $p_n$ are evolved using the
relations \eqref{gal_plus} and \eqref{gal_min} and we always have
$m_n+e_n+p_n=1$.
Following along the same lines as the analysis of the original SC decoder in 
\cite{Ari09}, we first claim that as $n$ grows large, the process $D_n$ will become polarized,
 i.e., the output of the
 process $D_n$ will almost surely be a completely noiseless or a completely erasure channel.
\begin{lemma}\label{polarize}
The random sequence $\{D_n=(p_n,e_n,m_n), n \geq 0 \}$ converges almost surely to a random 
variable $D_\infty$ such that $D_\infty$ takes its
value in the set $\{(1,0,0),(0,1,0)\}$.    
\end{lemma}
We now aim to compute the value of $C(W,Q)=\text{Pr}(D_\infty = (1,0,0))$, i.e., the highest rate
that we can achieve with the 1-Bit Decoder with Erasures. In this regard, 
a convenient approach is to find a function $f: \mathcal{D} \to \mathbb{R}$ 
such that $f((0,1,0))=0$ and $f(1,0,0)=1$ and for any $D \in \mathcal{D}$ 
\begin{equation*}
 \frac12 (f(D^+) + f(D^-))=f(D). 
\end{equation*}
With such a function $f$, the process $\{f(D_n)\}_{n \geq 0 }$ is a martingale and 
consequently we have $\text{Pr}(D_\infty = (1,0,0))=f(D_0)$. Therefore, by computing 
the deterministic quantity $f(D_0)$ we obtain the value of $C(W,Q)$. However, finding a 
closed form for such a function seems to be a difficult task\footnote{The function $f$ clearly exists as one trivial candidate for it is $f(D)= \text{Pr}(D_\infty = (1,0,0))$, where $D_\infty$ is the limiting r.v. that
 the process $\{D_n\}_{n \geq 0}$ with starting value $D_0=D$ converges to.}. Instead, the idea is to look for alternative functions, 
denoted by $g: \mathcal{D} \to \mathbb{R}$, such that the process $g(D_n)$ is a super-martingale (sub-martingale) and hence
we can get a sequence of upper (lower) bounds on the value of $\text{Pr}(D_\infty = (1,0,0))$ as follows.
Assume we have a function $g: \mathcal{D} \to \mathbb{R}$ such that $g((0,1,0))=0$ and $g(1,0,0)=1$
and for any $D \in \mathcal{D}$,   
\begin{equation} \label{g}
 \frac12 (g(D^+) + g(D^-)) \leq g(D). 
\end{equation} 
Then, the process $\{g(D_n)\}_{n\geq 0}$ is a super-martingale and for $n \geq 0$ we have 
\begin{equation}
 \text{Pr}(D_\infty = (1,0,0)) \leq \mathbb{E} [g(D_n)].
\end{equation}
The quantity $\mathbb{E} [g(D_n)]$ decreases by $n$ and by using Lemma~\ref{polarize} we have
\begin{equation}
\text{Pr}(D_\infty = (1,0,0))= \lim_{n \to \infty}  \mathbb{E} [g(D_n)].
\end{equation}
In a similar way, on can search for a function  $h : \mathcal{D} \to \mathbb{R} $ such that for $h$
with the same properties as $g$ except that the inequality \eqref{g} holds in opposite direction and
in a similar way this leads us to computable lower 
bounds on $C(W,Q)$. It remain to find some suitable candidates for $g$ and $h$. 
Let us first note that 
a density $D$ as in 
\eqref{gal_den} can be equivalently represented as 
a simple BMS channel given in Fig.~\ref{fig:channel}. 
This equivalence stems from the fact 
that for such a channel, conditioned on the event that the 
symbol $+1$ has 
been sent, the distribution of the output is precisely $D$. With a 
slight abuse of notation, we also denote the corresponding BMS channel by $D$. In 
particular, it is an easy exercise to show that the capacity ($I(D)$),
the Bhattacharyya parameter ($Z(D)$) and the error probability ($E(D)$) of the density $D$ are given  as  
\begin{align*}
& I(D)= (m+p)(1- h_2(\frac{p}{p+m})),\\
& Z(D) = 2 \sqrt{mp}+e,  E(D)=1-p-\frac{e}{2},
\end{align*}
where $h_2(\cdot)$ denotes the binary entropy function.
Since the function $Q$ is a not an injective function, we have 
$ \frac{I(D^+)+ I(D^-)}{2} \leq I(D)$.
This implies that the process $I_n=I(D_n)$ is a bounded supermartingale. Furthermore, since $I(D=(1,0,0))=1$ and $I(D=(0,1,0))=0$, 
we deduce from Lemma~\ref{polarize} that $I_{n}$ converges a.s. to a $0-1$ valued r.v.  $I_{\infty}$ and
hence 
\begin{equation*} \label{nn}
C(W,Q)=\text{Pr}(D_\infty=(1,0,0))=\text{Pr}(I_\infty=1)=\mathbb{E}(I_\infty).  
\end{equation*}
Now, from the fact that $I_n$ is a supermartingale, we obtain  
\begin{equation}\label{U_n}
C(W,Q)\leq \mathbb{E}[I_n]\triangleq U_n,
\end{equation}
for $n \in \naturals$. In a similar way, one can obtain a sequence of lower bounds for $C(W,Q)$.
\begin{lemma} \label{h(D)}
 Define the function $F(D)$ as $F(D)=p-4\sqrt{pm}$ for $D \in \mathcal{D}$. We have 
$F(D=(1,0,0))=1$, $F(D=(0,1,0))=0$ and 
$\frac{F(D^+)+ F(D^-)}{2} \geq F(D)$.
\end{lemma}
Hence, the process $F_n=F(D_n)$ is a submartingale and 
for $n\in \naturals$ we have 
\begin{equation}\label{L_n}
 C(W,Q)\geq \mathbb{E}[F_n]\triangleq L_n. 
\end{equation}
Given a BMS channel $W$, one can numerically 
compute $C(W,Q)$ with arbitrary accuracy  
using the sequences $L_n$ and $U_n$
(see Figure~\ref{fig:dwecap}).  Also, for a channel $W$ with capacity $I(W)$ and error
 probability $E(W)$, we have
\begin{equation}
E(W) \leq \frac{1-I(W)}{2}.
\end{equation}
 Therefore, $\displaystyle \inf_{\{D: E(D)=\frac{1-I(W)}{2}\}} C(D,Q) \leq C(W,Q)$, which
 leads to the universal lower bound obtained in Figure~\ref{fig:dwecap}.
\begin{example}
 Let the channel $W$ be a BSC channel with cross over probability $\epsilon=0.11$ (hence $I(W) \approx 0.5$). Using \eqref{D_0} we obtain
\begin{equation}   \label{D_0}
D_0=  \left\{   
\begin{array}{lr}
\infty, &   \text{w.p. $ 1-\epsilon=0.89$},\\   
-\infty, &   \text{w.p.
$\epsilon=0.11$}. \\   
\end{array} \right.    
\end{equation}
Therefore, we get $L_0=F(D_0)=-0.361$ and $U_0=I(D_0)=0.5$. We can also compute $L_1=\frac{F(D_0^+)+F(D_0^-)}{2}=-0.191$,  
$U_1=\frac{I(D_0^+)+I(D_0^-)}{2}=.5$ and
\begin{align*}
& L_2=\frac{F(D_0^{++})+F(D_0^{+-})+F(D_0^{-+})+F(D_0^{--})}{4}=-0.075,\\
& U_2=\frac{I(D_0^{++})+I(D_0^{+-})+I(D_0^{-+})+I(D_0^{--})}{4}=0.498.
\end{align*}
Continuing this way, one can find $L_{10}=0.264, U_{10}=0.474$ and $L_{20}=0.398, U_{20}=0.465$ and so on. \EEx
% \begin{table}
% \centering
% \begin{tabular}{c c c c c c c }
% $n$ & $10$  & $20$  & $30$    & $40$   \\
% \hline
% $L_n$ & $0.264$  & $0.368$  & $0.398$  & $0.406$  \\
% \hline
% $U_n$ &$0.474$  & $0.466$  & $0.465$  & $0.464$  \\
% \end{tabular}
% \caption{ }
% \label{bounds}
% \end{table}  
 
\end{example}

\subsubsection{Scaling behavior and error exponent}
In the last step, we need to show that for rates below $C(W,Q)$ the block-error probability decays to $0$ for large block-lengths.
\begin{lemma} \label{exponent}
 Let $D \in \mathcal{D}$. We have
\begin{align*}
& Z(D^-) \leq 2Z(D) \text{ and }  Z(D^+) \leq 2(Z(D))^{\frac 32}. \label{plus_bound}
\end{align*}
Hence, for transmission rate $R < C(W,Q)$ and block-length $N=2^n$, the probability
 of error of $\text{SCD}_Q$, denoted by $P_{e,Q}(N,R)$ satisfies $P_{e,Q}(N,R)=o(2^{-N^{\beta}})$ for $\beta < \frac {\log \frac 32}{2}$.
\end{lemma}

% 
% 
% \section{Conclusion and Open Problems} \label{sec:conclusion}
% We have shown that polar codes are very robust with respect to
% quantization at the decoder -- even very simple decoders with only
% a few messages achieve a high fraction of the capacity.  This is
% good news if we are interested in a low-complexity implementation.
% 
% Not all news is good. Numerical calculations indicate that the speed
% of the polarization is in general further decreased by quantization.
% This means that we need to construct even longer codes.
% 
% A precise characterization of this trade-off, namely the trade-off between
% the polarization speed and the quantization would be of considerable practical
% value.


\section*{Acknowledgment}
The authors wish to thank anonymous reviewers for their
valuable comments on an earlier version of this
manuscript.  The work of Hamed Hassani was supported by Swiss National 
Science Foundation Grant no 200021-121903.



%\bibliographystyle{IEEEtran}

\end{document}